\documentclass[intlimits,twoside,a4paper]{article}

\usepackage[T2A,T1]{fontenc} 

\usepackage{cmpj3}

\usepackage{multirow}

\issue{2025}{28}{4}{43706}
\doinumber{10.5488/CMP.28.43706}

\title[\textit{Ab initio} study of mechanical and functional properties of novel CaZnC and CaZnSi half-Heusler materials]
{Ab initio study of mechanical and functional properties of novel CaZnC and CaZnSi half-Heusler materials} 
\author[P. K. Kamlesh, U. K. Gupta, S. Verma, M. Rani, Y. Toual, A. S. Verma] {%
P. K. Kamlesh\orcidlink{0000-0001-7361-3519}\refaddr{label1}\thanks{Corresponding author: peeyush.physik@gmail.com.}, %
U. K. Gupta\orcid{0000-0002-9879-9150}\refaddr{label2}, %
S. Verma\orcid{0000-0003-4962-6226}\refaddr{label1}, %
M. Rani\orcid{0000-0002-8402-4522}\refaddr{label3}, %
Y.~Toual\orcid{0000-0002-6439-8527}\refaddr{label4}, %
A.~S.~Verma\orcid{0000-0001-8223-7658}\refaddr{label5,label6}}

\addresses{
\addr{label1} Department of Physics, Poornima University, Jaipur 303905, Rajasthan, India
\addr{label2} School of Electrical \& Power Engineering, Anand International College of Engineering Kanota, Jaipur 303012, Rajasthan, India
\addr{label3} Department of Physics, Mohanlal Sukhadia University, Udaipur 313001, Rajasthan, India
\addr{label4} Laboratory of Solid Physics, Sidi Mohamed Ben Abdellah University, Faculty of Sciences, BP 1796, Fez, Morocco
\addr{label5} Department of Physics, Anand School of Engineering \& Technology, Sharda University Agra, Keetham, Agra 282007, India
\addr{label6} University Center for Research and Development, Department of Physics, Chandigarh University, Mohali, Punjab 140413, India
}

\Keywords{optical properties, structural stability, thermodynamic stability, half-Heusler, thermoelectric materials}

\date{Received August 22, 2025, in final form November 24, 2025}

\begin{document}

\maketitle

\begin{abstract}
This research work introduces the DFT through FP-LAPW+lo technique in WIEN2k software to obtain information about structural, thermoelectric, and optoelectronic characteristics of CaZnC and CaZnSi materials. The structural optimization was performed using PBE-GGA functional, while the rest of the characteristics were obtained with the PBE-GGA + TB-mBJ approach. The thermoelectric parameters were evaluated using BoltzTraP software. The elastic constants and other mechanical parameters were computed by utilizing the ELAST code within the WIEN2k software, while the thermodynamic characteristics were evaluated using the Gibbs2 program. The findings show a correlation between atomic composition and lattice dimensions while finding that CaZnC has a direct ($\Gamma$--$\Gamma$) band gap of $1.186$~eV, whereas CaZnSi has an indirect ($\Gamma$--$X$) band gap of $1.067$~eV. The optical studies of the compounds show potential applications for photovoltaics while the thermoelectric results find optimized power factors and figure of merit values for energy conversion performance. The elastic parameters of CaZnC and CaZnSi demonstrate material stability and brittleness. Lastly, the thermodynamic evaluations provide information about the thermal mechanism and disorder of the materials. As a result, this research work provides significant advancements in the understanding of the fundamentals of these compounds and highlights their promising applications in renewable energy technologies.
\printkeywords

\end{abstract}

\section{Introduction}

In recent times, several researchers have directed their efforts towards exploring smart and versatile materials capable of serving diverse purposes such as sensors, optoelectronic devices, spintronic technologies, thermoelectric (TE) systems, and various applications in the renewable energy field~\cite{ref1,ref2,ref3,ref4}. The escalating rate of carbon emissions has become a pressing global concern, posing a significant threat to the environment. As a result, there is more need and urgency to find clean, truly renewable energy sources, such as TE, bio-fuel, bio-mass, solar energy, geothermal energy, and wind energy. To solve the environmental issues, there is an urgent need for materials that serve in both photovoltaics and TE generators. Photovoltaics can harness the large amounts of solar energy constantly expelled from the sun into useful electric energy, while TE generators represent a way to take advantage of dissipated energy in order to promote energy sustainability. Together, these two energy generation technologies fulfill the need for green energy resources.

Recent studies have identified materials for both photovoltaic and TE applications, including chalcogenides, perovskites, organic materials, and alloys~\cite{ref5,ref6,ref7,ref8,ref9,ref10}. Half-Heusler (HH) alloys are a very promising class of materials and receive tremendous interest for their numerous useful properties such as high melting points, magnetic properties, and TE properties~\cite{ref10,ref11,ref12,ref13}. HH materials have other specific properties, such as piezoelectric and optoelectronic, which may also be useful for energy applications. In addition to their potential use in TE applications, HH materials can also be applied to many advanced applications. Overall, they represent exciting materials in the rapid development of effective heat transfer applications, superconductors, spintronic, high-temperature phononic devices, shielding from ultraviolet (UV) radiation, solar energy, optical laser diodes~\cite{ref14,ref15,ref16,ref17,ref18}.

From the initial identification of Heusler materials by Friedrich Heusler~\cite{ref19} in 1903, these materials have developed to occupy a large space within the scientific community. HH materials crystallize in a cubic arrangement ($F\bar{4}3m$ space group) and have unique properties. HH materials can be categorised into two groups based on their Valence Electron Count (VEC), where some materials have a VEC of 8 and others have a VEC of 18. This categorization scheme highlights different structural and electronic properties of the HH family, and makes their potential application to other areas of science more pronounced. TE generators operate based on the Seebeck effect which generates an electrical voltage in a material, where there is a temperature difference across that material. However, TE devices have, in most cases, a very low efficiency in keeping with their figure of merit (ZT). It is derived from the equation:

\begin{equation}
ZT = \frac{S^2 \, \sigma T}{\kappa_{\text{total}}}, \label{1}
\end{equation}
where, $S$ denotes the Seebeck parameter, $\sigma$ denotes the electrical conductivity, $T$ denotes the absolute temperature, and $\kappa_{\text{total}}$ depicts total thermal conductivity (electron + phonon).

Therefore, materials exhibiting low thermal conductivity along with elevated power factors are regarded as optimal candidates for improving the efficiency of TE equipment. By minimizing thermal conductivity and maximizing power factors, these compounds hold promise in improving the efficiency of TE generators~\cite{ref20, ref21}.

Gruhn~\cite{ref22} studied tailored semiconductor materials for optoelectronic applications, specifically thin-film photovoltaic cells and optical laser diodes. He studied 648 ternary 1:1:1 HH materials through \textit{ab initio} calculations and found preferred stacking configurations and trends in the behavior of semi-conductivity. His calculations of the lattice geometries allowed him to calculate estimates for lattice constants. He also used calculated bandgaps to identify substitutes for CdS in shield layers of thin-film solar cells. Belmiloud~et al. \cite{ref23} worked on the structural and electronic parameters of new HH materials as a systematic study to use as potential photovoltaic applications. They premeditated five ternary materials (NaAgO, CaZnC, ScAgC, YCuC, and LiCuS) using density functional theory (DFT) which were found to be thermodynamically stable and to have direct energy gaps close to 1~eV. The compounds exhibited quasi-lattice matching with GaAs (Si) and had mechanical and dynamical stabilities, which is positive for technologies of solar cells. Azouaoui~et al. \cite{ref24} studied the structural details as well as other physical parameters of NaCaA (A= As, P, N) HH semiconductors up to 20~GPa with DFT. Their results indicate that these compounds are chemically stable in the $\alpha$-phase structure and semiconductors with indirect bandgaps and good optical properties. NaCaZ materials are mechanically stable, some of which behave ductilely above 10~GPa. The phonon dispersion calculations reveal that the materials are dynamically stable at pressure. These results highlight the prospects of NaCaZ materials for several applications, including optoelectronics and solar cell absorbers. Xiong et al. \cite{ref25} carried out an investigation into the TE properties of HH compounds LiAlTt (Tt $=$ Si, Ge) with eight valence electrons. Both LiAlGe and LiAlSi displayed elevated Seebeck coefficients around room temperature. LiAlSi demonstrated $n$-type behavior, while LiAlGe exhibited $p$-type characteristics. Although LiAlGe has high thermal conductivity, incorporation of LiAlSiGe solid solutions substantially enhanced its TE performance, yielding a $ZT$ four times greater than previous findings for LiAlGe. This reveals the potential of 8-VEC HH materials in TE applications. Wu et al. \cite{ref26} applied first-principles calculations to study how pressure impacts on the dynamical and mechanical stability of NaAlSi. They observed a decline in lattice constants and a rise in elastic constants under pressure, with mechanical instability occurring when $C_{44}$ decreases above 25.97~GPa. Phonon-dispersion curves suggested structural stability at ambient pressure but predicted instability beyond 27.52~GPa, indicating a potential phase transition. Jin et al. \cite{ref27} explored the topological energy bandstructure of NaAlSi material, revealing four type-I nodal lines near to Fermi level and the presence of drumhead surface states. Yi et al. \cite{ref28} proposed NaAlSi(Ge) as dual double node-line semi-metals exhibiting distinctive surface states, suitable for studying correlated phases and as cathode materials used in the cathodes of sodium-ion batteries. Ciftci's \cite{ref29} \textit{ab initio} study confirmed NaAlSi's mechanical stability as a semiconductor with potential in optoelectronic applications. Wang  et al. \cite{ref30} investigated NaAlGe, identifying it as a topological nodal line semimetal with distinct surface states, suitable for electronic devices. These studies collectively are the showcase of the diverse properties and potential applications of these materials.

The literature review underscores a notable lack of investigation into Ca-based HH-materials having a VEC of 8. Moreover, there is a research gap in the exploration of both optical and TE characteristics within CaZnC and CaZnSi HH-compounds. This breakthrough serves as a major catalyst for additional research and calls for an in-depth investigation into the fundamental properties of CaZnC and CaZnSi HH compounds. The discussion will include an assessment of structural, mechanical and thermal stabilities, as well as electronic, optical, and TE properties. Taken holistically; the evaluation should enable the assessment of the opportunities for introducing this material into renewable sectors, to add functionality and efficiencies in sustainable use. To our knowledge as limited evaluation of their optical properties and TE behaviour has occurred in the published literature. Therefore, this study would address this limitation, to facilitate a complete understanding of the endeavoured applications in photovoltaics linked with TEs. This would increase the possibilities of exploiting their novel properties for next generation and enhancement of energy harvesting and exploiting renewable energy developments into sustainable applications with greater efficiency.

\section{Research methodology}

The first step involved conducting structural stability analysis utilizing DFT model with the FP-LAPW+lo method in WIEN2k software \cite{ref31,ref32,ref33}. This method ensures accurate predictions of atomic positions and energy states. Next, we analyzed the vibrational properties and electronic structures to assess the material's stability and conductivity. 
For present calculations, a cutoff energy of $-6.0$~Ry was used/choosen. Each atoms muffin-tin (MT) radius was adjusted to stop the charge leakage.
Furthermore, the spherical harmonics for atomic sphere were adjusted to $l_{\text{max}} = 10$ to ensure a comprehensive representation of the electronic states. In this work, $R_{\text{MT}}K_{\text{max}}$ was set to 7.0 for both compounds; $K_{\text{max}}$ represents the maximum amplitude of reciprocal vector in first Brillouin zone (BZ), while $R_{\text{MT}}$ represents the shortest MT radius in the unit cell (u.c.). Moreover, $G_{\text{max}}$ corresponds to the size of the first vector in Fourier series representation of charge density. Here, a value assigned to $G_{\text{max}}$ is 12. A threshold of convergence for charge was taken $10^{-3}e$ while for energy it was taken $10^{-4}$~Ry to ensure the convergence of charge and energy alterations for each next SCF cycle. Along the first BZ, a $10 \times 10 \times 10$ $k$-point mesh was applied for obtaining fundamental properties and stability criteria.

The structural optimization and electronic parameters were computed through the ``generalized gradient approximation'' (GGA) in the framework of PBE \cite{ref34}. To enhance the bandgap, the ``Tran-Blaha modified Becke-Johnson'' (TB-mBJ) \cite{ref35} approach was used in conjunction with the PBE-GGA method. This choice was made because the PBE-GGA method has a tendency to underestimate the bandgap.

The TE characteristics were evaluated by applying the BoltzTraP code \cite{ref36} incorporated in WIEN2k. The code investigates transport characteristics under the static relaxation time method ($\tau$). It is formulated by the semi-classical Boltzmann transport equation. The ELAST package was utilized to evaluate the elastic constants. The package is executed within WIEN2k simulation code. The approach is founded on investigating the ground state of total energy, which offers a suitable way to extract elastic constants. Other mechanical properties were also evaluated. The evaluation used the Voigt-Reuss-Hill (VRH) approach. The VRH approach was used with the same package. The thermodynamic parameters were evaluated by utilizing the Gibbs2 \cite{ref37} program. The evaluation was completed for the temperature range of $0$~K to $1000$~K and the pressure range from $0$~GPa to $10$~GPa.

\section{Results and discussion}

\subsection{Structural parameters}

\begin{figure}[!t]
\centering
\includegraphics[width=0.4\textwidth]{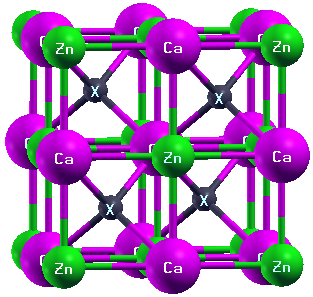}
\caption{(Colour online) Unit cell and structure of CaZnX ($X =$ C and Si) HH compounds.}
\label{fig:1}
\end{figure}

The CaZnC and CaZnSi HH compounds are synthesized in a cubic system in $F\bar{4}3m$ (\#216) space group. The position of atoms can be specified by Wyckoff positions as: the Ca atoms occupy Wyckoff positions of type $4b$ at $(0.5,\,0.5,\,0.5)$ and the Zn atoms occupy Wyckoff positions of type $4a$ at $(0,\,0,\,0)$. The Wyckoff positions of type $4c$ at $(0.25,\,0.25,\,0.25)$ are occupied by $X$, i.e., C or Si atom. The crystalline structure is a crucial aspect in governing the physical and electronic properties of a substance. The unit cell structures of these investigated materials are illustrated in figure~\ref{fig:1}.  

Energy minimization calculations were performed to confirm the most stable ground state of inspected compounds. Subsequently, variations in energy with varying volume were analyzed, and the data were fitted to Birch-Murnaghan's equation of state \cite{ref38}:

\begin{equation}
E(V) = E_{0} + \frac{9V_{0}B_{0}}{16} \left\{ \left[ \left(\frac{V_{0}}{V}\right)^{2/3} - 1 \right]^{3} B_{0}^{\prime} + \left[ \left(\frac{V_{0}}{V}\right)^{2/3} - 1 \right]^{2}  \left[6 - 4\left(\frac{V_{0}}{V}\right)^{2/3}\right] \right\}. 
\tag{2}
\label{eq:birch}
\end{equation}

\begin{figure}[!t]
    \centering
        \centering
       \includegraphics[width=0.48\textwidth]{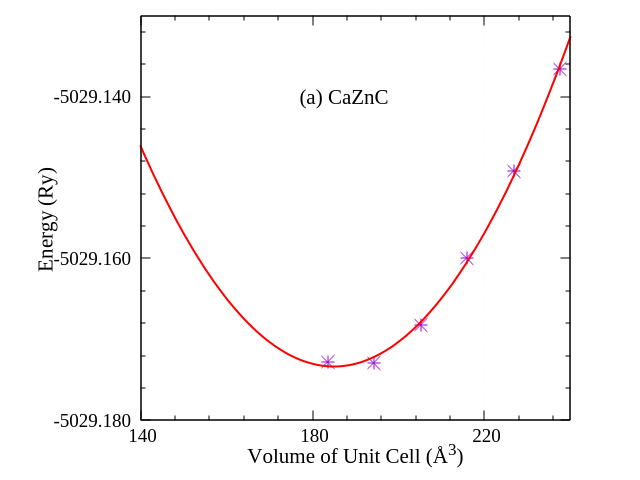}
        \centering
        \includegraphics[width=0.48\textwidth]{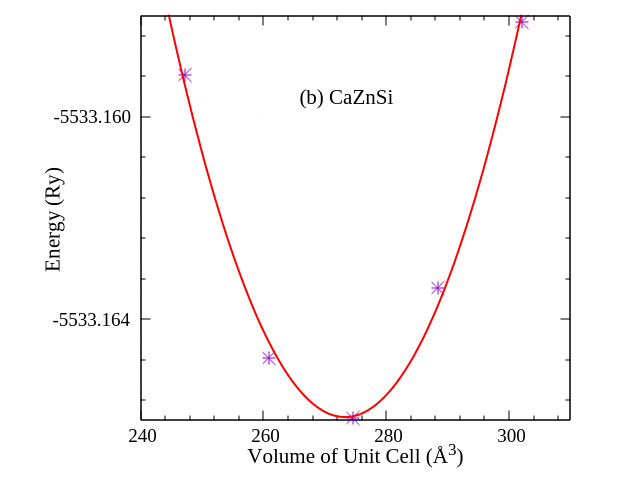}
   \caption{(Colour online) Volume optimization plot for the HH compounds (a) CaZnC and (b) CaZnSi.}
    \label{fig:2}
\end{figure}

Here, $V$ represents the volume at normal pressure, while $V_{0}$ represents volume at zero external pressure. Figure~\ref{fig:2} presents energy-volume curves for the investigated substances. Through this analysis, various structural parameters were determined. These include the equilibrium energy ($E$ in Ry) and atomic lattice parameter ($a$ in \AA).The zero-pressure bulk modulus, ($B_{0}$ in GPa), and its first pressure derivative ($B_{0}^{\prime}$) were computed.

Table~\ref{tab:1} displays the obtained structural properties. It is evident that elevated atomic numbers of the $X$-atom are associated with enhanced values of the lattice parameters. This relationship suggests a correlation between the atomic composition and the overall structural dimensions of the materials. Another crucial structural parameter examined is the bulk modulus, which assesses the stability of a compound when subjected to volume variations under pressure. Interestingly, the bulk modulus follows an inverse relationship with the lattice constant. At absolute pressure, it decreases as atomic number of $X$ atom increases.  

\begin{table}[!t]
\centering
\caption{Structural, electronic, and optical characteristics of CaZnC and CaZnSi HH materials.}
\vspace{0.2cm}
\label{tab:1}
\resizebox{\textwidth}{!}{%
\begin{tabular}{lcccccccccc}
\hline
Compounds & $a$ (\AA) & $B_{0}$ (GPa) & $B_{0}^{\prime}$ & $E$ (Ry) & Bandgap PBE (eV) & Bandgap PBE+mBJ (eV) & $\varepsilon_{1}(0)$ & $n(0)$ & $R(0)$ \\
\hline
CaZnC     & 5.739 & 78.928 & 4.543 & $-5029.174$ & 0.329 & 1.186 & 8.722 & 2.953 & 0.244 \\
          & 5.740$^{a}$ & 76.2$^{a}$ &  --   & --        & --    & 1.177$^{a}$ & -- & -- & -- \\
CaZnSi    & 6.475 & 47.140 & 3.655 & $-5533.166$ & 0 & 1.067 & 13.704 & 3.702 & 0.330 \\
CaMgSi$^{b}$ & 6.347$^{b}$ & 43.68$^{b}$ & 3.52$^{b}$ & -- & 0.709$^{b}$ & 1.315$^{b}$ & -- & -- & -- \\
\hline
\multicolumn{10}{l}{$^{a}$Reference~\cite{ref23}; $^{b}$Reference~\cite{ref39}} \\
\end{tabular}}
\end{table}

This finding underscores the relationship between the bulk modulus as well as compressibility, which are inversely proportional to one another. Consequently, the compressibility varies as CaZnC $<$ CaZnSi. Moreover, the bulk modulus pressure derivative provides important insights into a compound's thermoplastic behavior. It indicates the material's reaction to mechanical stress. It also reflects its deformation under varying pressure conditions. For every material examined in this study, a positive bulk modulus pressure derivative was observed. This suggests that these substances tend to become more rigid with increasing pressure.

\subsection{Electronic parameters}

\begin{figure}[!t]
    \centering
        \centering
        \includegraphics[width=5cm, height=8cm]{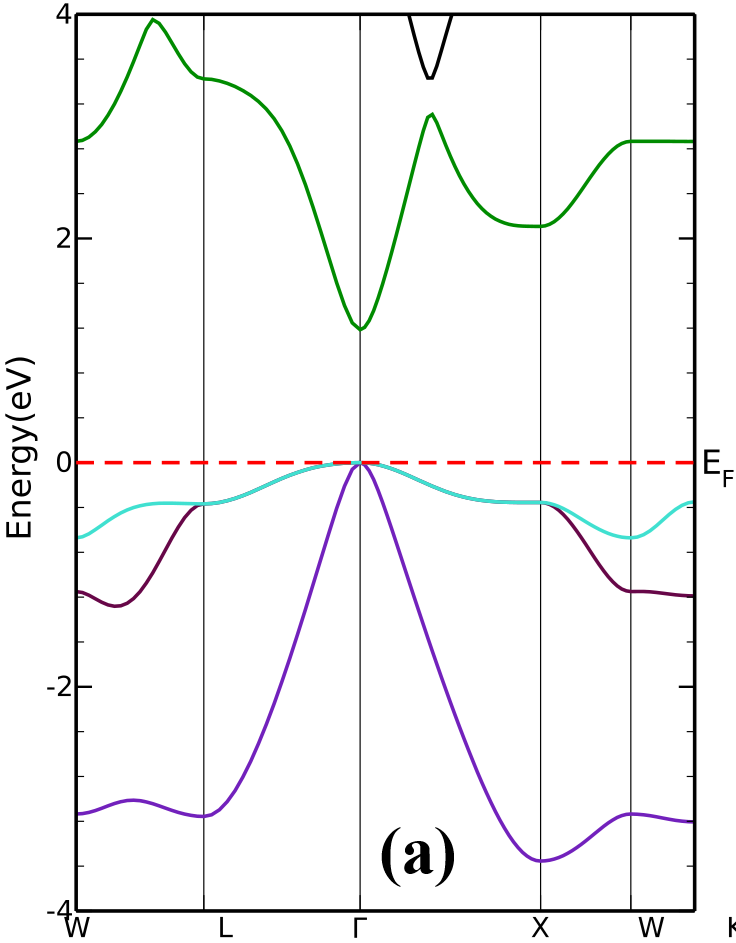}
        \centering
        \includegraphics[width=5cm, height=8cm]{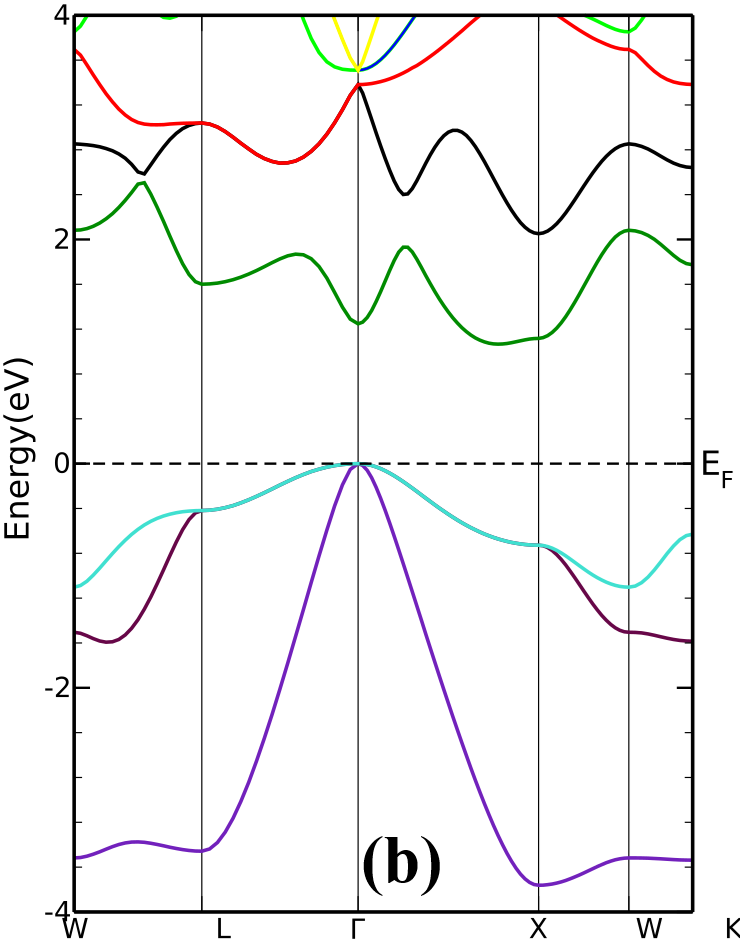}
   \caption{(Colour online) Bandstructure curves of (a) CaZnC and (b) CaZnSi HH materials, calculated using the TB-mBJ potential.}
    \label{fig:3}
\end{figure}

Various physical characteristics of a compound are intricately linked to its electronic parameters. Achieving precision in the bandstructure is crucial as it serves as the foundation for determining other physical characteristics. Therefore, the authors first examined the bandstructure of these materials. Then they proceeded to investigate additional parameters. Initially, SCF computations were performed utilizing the PBE-GGA potential. The optimized lattice characteristics were applied in these calculations. CaZnC was found to have a low bandgap. In disparity, CaZnSi exhibited zero bandgap. This phenomenon may be due to the bandgap underestimation by the PBE-GGA potential, as noted in earlier studies~\cite{ref40,ref41}. To address this, the researchers employed TB-mBJ method to modify outcomes obtained from PBE-GGA potential functional. This method modifies specific input components of the semi-local potentials with the objective of approximating bandgaps that are more consistent with experimental observations~\cite{ref42}. After applying TB-mBJ, the energy gap of CaZnC and CaZnSi found to be 1.186~eV and 1.067~eV, correspondingly (table~\ref{tab:1}) showing their semiconductor behavior. Figure~\ref{fig:3} presents the bandstructure diagrams of these compounds under standard pressure conditions, highlighting how energy correlates with the wave vector $k$ within the first BZ. It is worth mentioning that Fermi level was assigned a value of zero eV for both materials. Examination of figures~\ref{fig:3}(a) and \ref{fig:3}(b) reveals a distinct feature: CaZnC demonstrates a direct ($\Gamma$--$\Gamma$) bandgap, whereas CaZnSi shows an indirect ($\Gamma$--$X$) bandgap. Further examination of bandstructures reveals intriguing details. For both compounds, degeneracy occurs at $\Gamma$, $L$, and $X$ points. This occurrence of degenerate electronic bands indicates a covalent bond between the constituent atoms of the compounds studied.

\begin{figure}[!t]
    \centering
    \includegraphics[width=12cm, height=8cm]{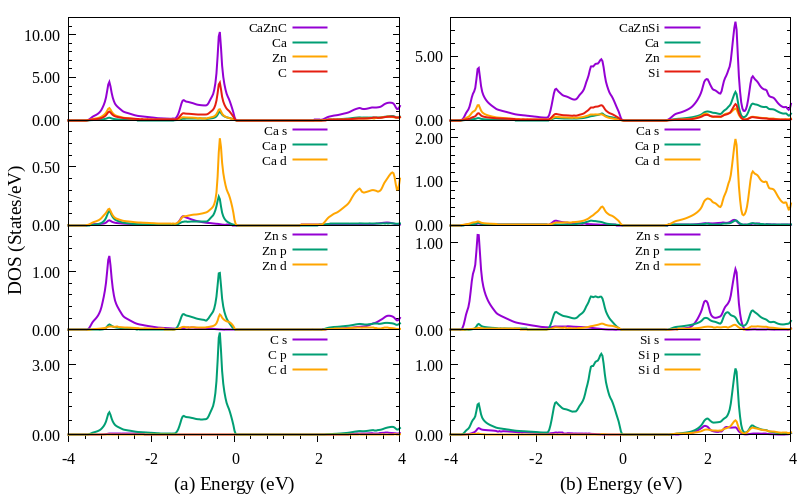} 
    \caption{(Colour online) Total and partial DoS plots of (a) CaZnC and (b) CaZnSi HH materials.}
    \label{fig:4}
\end{figure}

To gain a deeper insight into the interactions between different atomic orbitals, the total and partial density of states (DOS) of the materials were investigated by utilizing a combination of PBE-GGA potential with TB-mBJ technique. The results are depicted in figure~\ref{fig:4}, revealing a striking similarity in the computed bandstructure and DOS for both materials under study.  

A detailed examination of the lower valence band ranging $-4.0$~eV to $-2.0$~eV indicates that the $s$-orbitals of Zn atoms contribute more significantly compared to the $p$-states of the C/Si atoms. By contrast, the electronic contribution from Ca atoms in this region is minimal. As we go into the higher valence band region ($-2.0$~eV to $0.0$~eV), the significant input appears from the $p$-orbitals  of C/Si atoms while the involvement of $d$-state of Ca and $p$-state of Zn is relatively much less.

In the conduction band, however, the main contribution arises from the $d$-orbital of the Ca atom while lesser involvement comes from the $p$-states of Zn and C/Si atoms. Such a full examination of the orbital contributions permits a more specific perception of the electronic structure and interatomic interaction for the addressed compounds, which means their intrinsic properties and related applications.

\subsection{Optical parameters}

The optical behaviours of any compound are fundamental in anticipating the interaction of electromagnetic (EM) radiation with the compound, thereby determining the potential applications of the compound into optoelectronics. In this study, a number of optical properties are investigated such as reflectivity $R(\omega)$, refractive index $n(\omega)$, optical conductivity $\sigma(\omega)$, extinction coefficient $k(\omega)$, dielectric constant $\varepsilon(\omega)$, absorption coefficient $\alpha(\omega)$, and energy loss function $E_{\text{loss}}(\omega)$.

Among these values, the dielectric constant $\varepsilon(\omega)$ for one compound explains the interaction between its electrons and photons, basically defining the response of the compound to the incident EM radiation under an infinitesimal wave vector, hence giving insight into its optical behavior and electronic polarization characteristics. This function is typically represented by Ehrenreich and Cohen's equation~\cite{ref43} and separates into two components:  

\begin{equation}
\varepsilon(\omega) = \varepsilon_1(\omega) + i\varepsilon_2(\omega). \tag{3}
\end{equation}

In this context, $\varepsilon_1(\omega)$ denotes the real fraction of the dielectric tensor, while $\varepsilon_2(\omega)$ represents its imaginary counterpart. Consequently, it is apparent that the dielectric constant $\varepsilon(\omega)$ is a complex quantity, where $\omega$ corresponds to the angular frequency of EM energy. The real part, $\varepsilon_1(\omega)$, offers information regarding electronic polarization and anomalous dispersion within the material, the imaginary part, $\varepsilon_2(\omega)$, clarifies the absorption processes occurring within it. The formulation for $\varepsilon_2(\omega)$ can be derived as outlined in reference~\cite{ref44}:

\[
\varepsilon_2(\omega) = \frac{4 e^2 \piup^2}{\omega^2 m^2} 
\sum_{i,j} \int \langle i | M | j \rangle \, \nabla^2 f_i \, (1 - f_i) \,
\delta(E_f - E_i - \hbar \omega) \, \rd^3k. 
\tag{4}
\]

Here, $M$ represents the momentum operator, while $i$ and $j$ belong to the initial and final states in the valence and conduction bands, respectively. The term $f_i$ denotes the Fermi distribution function of the initial state and $\delta(E_f - E_i - \hbar\omega)$ represents the change in energy between initial and final states due to absorption of an incident photon with energy $\hbar\omega$. Additionally, $e$, $\omega$, $m$, and $\hbar$ correspond to the electron charge, angular frequency of the incident photon electron's mass, and reduced Planck constant, respectively.  By employing $\varepsilon_2(\omega)$, within the Kramers-Kronig relation~\cite{ref45}, $\varepsilon_1(\omega)$ can be computed as below:

\[
\varepsilon_1(\omega) = 1 + \frac{2}{\piup} 
\mathcal{P} \int_0^{\infty} 
\frac{\omega' \, \varepsilon_2(\omega')}{\omega'^2 - \omega^2} 
\, \rd\omega'.
\tag{5}
\]

Here, $\mathcal{P}$ indicates the principal value of the integral.

\begin{figure}[h!]
    \centering
     \includegraphics[width=12cm, height=8cm]{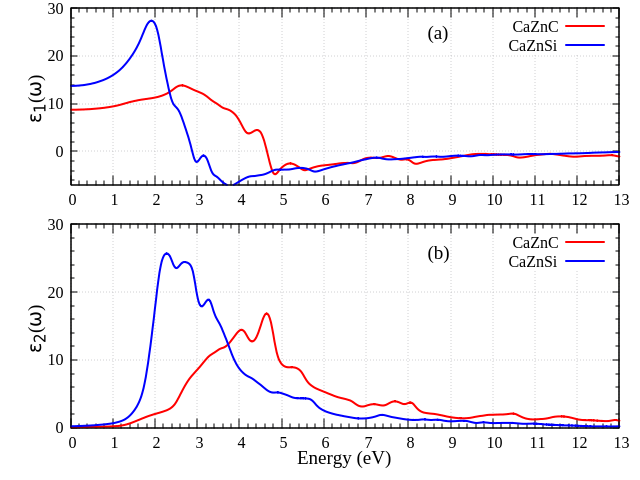} 
    \caption{(Colour online) Change in (a) $\varepsilon_1(\omega)$ and (b) $\varepsilon_2(\omega)$ of CaZnC and CaZnSi HH materials with incident radiation energy.}
     \label{fig:5}
\end{figure}

\begin{figure}[!t]
    \centering
   \includegraphics[width=12cm, height=8cm]{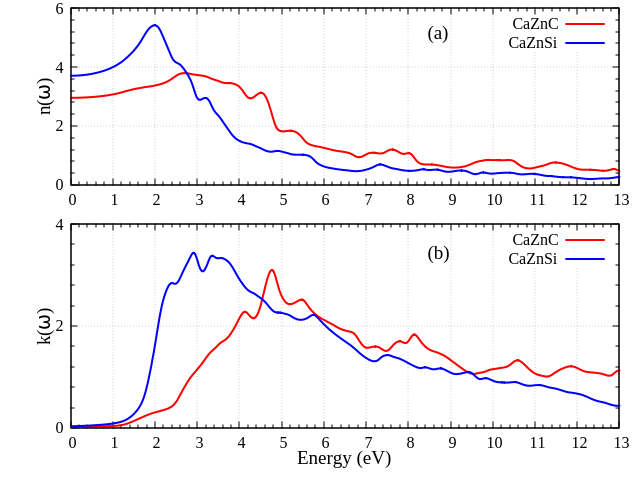}
    \caption{(Colour online) Change in (a) $n(\omega)$ and (b) $k(\omega)$ of CaZnC and CaZnSi HH materials with incident radiation energy.}
    \label{fig:6}
\end{figure}

\begin{figure}[!t]
    \centering
    \includegraphics[width=12cm, height=8cm]{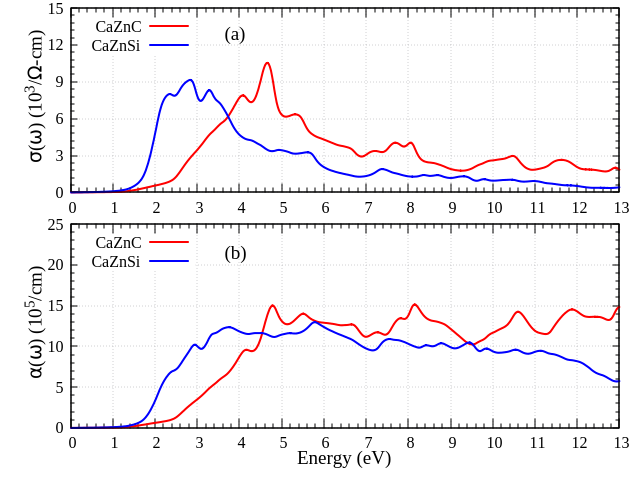}
    \caption{(Colour online) Change in (a) $\sigma(\omega)$ and (b) $\alpha(\omega)$ of CaZnC and CaZnSi HH materials with incident radiation energy.}
     \label{fig:7}
\end{figure}

\begin{figure}[!t]
    \centering
    \includegraphics[width=12cm, height=8cm]{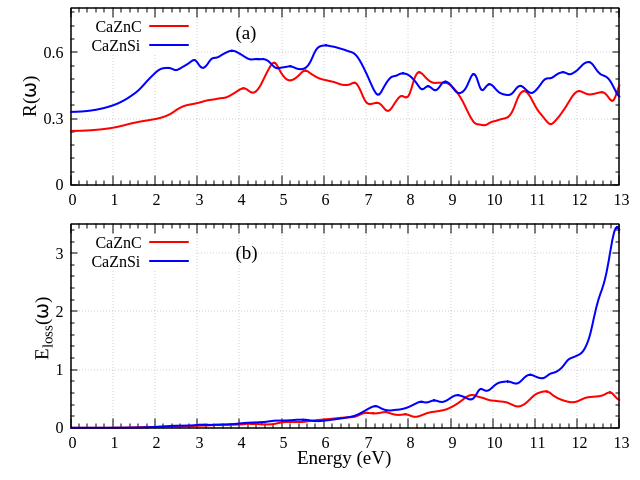} 
    \caption{(Colour online) Change in (a) $R(\omega)$ and (b) $E_{\text{loss}}(\omega)$ of CaZnC and CaZnSi HH materials with incident radiation energy.}
     \label{fig:8}
\end{figure}

To understand the behavior of the compound under consideration to incident energy, the dielectric tensor $\varepsilon(\omega)$ has been examined. Moreover, it is possible to infer other pertinent optical features from $\varepsilon_1(\omega)$ and $\varepsilon_2(\omega)$ \cite{ref46}. Figures (\ref{fig:5}--\ref{fig:8}) exhibit the optical components that have been estimated for the symmetry structure of the compounds. Spanning an energy range of incoming EM ray varies within a range of 0 to 13 eV. These calculations show that the materials are homogeneous and isotropic, meaning that their optical properties change depending on the frequency of the incident EM radiation but stay constant with different orientations of the electric field vector.

Figure~\ref{fig:5} shows the spectral variation of both, $\varepsilon_1(\omega)$ and $\varepsilon_2(\omega)$, with the energy of the incident radiation. The static dielectric constant, indicated by $\varepsilon_1(0)$, is determined within a narrow energy range and reflects the dielectric tensor of the material, which characterizes its response to the external electric field in absence of frequency-dependent effects. Specifically, for the CaZnC and CaZnSi compounds, the values of $\varepsilon_1(0)$ are 8.72 and 13.70, respectively. A greater value of $\varepsilon_1(0)$ signifies a more robust reaction of the substance to the applied EM energy. The fluctuation in $\varepsilon_1(\omega)$ with incident radiation for the materials under consideration is shown in figure~\ref{fig:5}(a). It is evident from these plots that $\varepsilon_1(\omega)$ rises with the growing energy of incident radiation and reaches a maximum at 2.63~eV and 1.92~eV for CaZnC and CaZnSi, respectively, then rapidly decreases within a small energy range and becomes negative at 4.67~eV and 2.90~eV, correspondingly. Following this, a small increase towards zero is obtained. Notably, the peak of $\varepsilon_1(\omega)$ for both materials is observed within the range of visible region. This indicates that these compounds exhibit maximal response in the visible region. 

The behavior of $\varepsilon_2(\omega)$ for the materials under investigation is depicted in figure~\ref{fig:5}(b), which displays the threshold energy of $\varepsilon_2(\omega)$ for CaZnC and CaZnSi as 1.18~eV and 0.86~eV, respectively. This figure represents the optical bandgap of a compound which agrees well with the determined electronic energy bandgap, confirming the validity of the results. Furthermore, $\varepsilon_2(\omega)$ exhibits distinct peaks, with the first peak originating from electronic transitions between valence band maxima and conduction band minima. The additional crests in $\varepsilon_2(\omega)$ arise due to the electron excitations between various energy states within the conduction and valence bands, highlighting absorption characteristics and interband transition dynamics of the material.

Figures~\ref{fig:6}(a) and \ref{fig:6}(b) show the variations in the refractive index $n(\omega)$ and extinction coefficient $k(\omega)$, respectively, for the materials studied. One of the basic optical parameters that determines how light travels through a material and how well it may absorb incident radiation, is its refractive index. The static refractive index, represented as $n(0)$, for CaZnC and CaZnSi substances, is determined to be 2.953 and 3.702, respectively. Upon examination of figures~\ref{fig:5}(a) and \ref{fig:6}(a), it becomes apparent that $n(\omega)$ exhibits a pattern closely resembling that of $\varepsilon_1(\omega)$ with an increase in incident radiation energy. Conversely, unlike $\varepsilon_1(\omega)$, $n(\omega)$ never becomes negative. This implies that even at high incident energy frequencies, the material retains its transparency. 

$k(\omega)$ adopts as a crucial parameter in describing the attenuation of the electric field oscillations of incident EM radiation within a material. A closer inspection of figure~\ref{fig:6}(b) reveals that $k(\omega)$ exhibits distinct peaks that closely correspond to the frequencies where $\varepsilon_1(\omega)$ touches its minimum or transitions to zero. This correlation indicates a strong relationship between optical absorption and the dielectric response of the material. After touching its peak value, $k(\omega)$ steadily decreases with an increase in incident radiation as energy, signifying a reduction in absorption at higher energy levels.

Optical conductivity, $\sigma(\omega)$, is an important property of a compound that describes how it conducts electrons when gas exposed to incident EM radiation. The $\sigma(\omega)$ variation with incident energy is shown in figure~\ref{fig:7}(a), which also portrays electronic excitation and charge transport behavior of a compound. $\sigma(\omega)$  for both CaZnC and CaZnSi increase as frequency of incident radiation increases, peaking before starting to drop off. Notably, the maximum optical conductivity for CaZnC occurs in the UV region at 4.67~eV, whereas for CaZnSi, the highest $\sigma(\omega)$ is detected in the visible region at 2.84~eV. Consequently, these compounds have great potential for use in photovoltaic devices. CaZnSi is more appropriate for devices in the visible region and CaZnC is more suitable for devices that operate in the UV region of the EM range. 

The absorption coefficient, $\alpha(\omega)$, quantifies the fraction of incident EM energy absorbed per unit thickness of a material. A higher $\alpha(\omega)$ value suggests an enhanced capability for electron excitation, facilitating the movement of electrons from the valence band to the conduction band. Figure~\ref{fig:7}(b) illustrates the change in $\alpha(\omega)$ for the investigated materials with varying incident radiation energy. This figure shows that conductivity seems to be at a minimum within the same region of incident energy where absorption reaches its minimum value. It is also interesting to note that incident radiation frequencies matching the absorption peaks align with optical conductivity peaks. This alignment validates theoretical expectations, affirming the precision of the computed results. 

Figure~\ref{fig:8}(a) illustrates the optical reflectivity, $R(\omega)$, of the examined substances. Static reflectivity, $R(0)$, for the studied materials, is measured to be 24.4\% and 33.0\% for CaZnC and CaZnSi respectively. Initially, the reflectivity $R(\omega)$ for both materials increases slowly and attains maximum value, followed by many fluctuations. The greatest reflectivity values for both compounds are found at 4.83~eV and 6.02~eV, respectively, in the UV spectrum of the incident EM radiation. Consequently, these materials demonstrate potential as effective shields against high-energy UV radiation. Moreover, they exhibit promise in PV relevance in the visible and low-energy UV regions. 

Figure~\ref{fig:8}(b) presents the energy loss function, $E_{\text{loss}}(\omega)$, of the examined compounds. This function describes how the energy of an electron decreases as it moves through a substance. The crests in the energy loss spectrum relate to the plasma resonance, a phenomenon arising from the combined oscillations of the valence band electrons. The specific frequency at which this resonance occurs is termed the plasma frequency ($\omega_P$). At this frequency, the material transitions from a dielectric to a metallic response, marking a significant shift in its optical behavior. The notable peaks of $E_{\text{loss}}(\omega)$ for CaZnC and CaZnSi are clearly located in the high-frequency UV range, at 11.25~eV to 12.94~eV, respectively, as can be seen in figure~\ref{fig:8}(b).

Table~\ref{tab:1} compiles the stationary values of the different optical components of the materials under investigation, such as the dielectric tensor $\varepsilon_1(0)$, refractive index $n(0)$, and optical reflectivity $R(0)$. Notably, $\varepsilon_1(0)$, $n(0)$, and $R(0)$ rise when C atom is replaced by Si atom in the considered HH compounds.

\subsection{Thermoelectric (TE) parameters}

The figure of merit ($ZT$) is a crucial factor in order to know the maximum achievable efficiency of TE substances. The $ZT$ is given in equation~\ref{1} , where $S^2\sigma$ denotes the power factor, encompassing the Seebeck coefficient $(S)$, electrical conductivity $(\sigma)$, and absolute temperature $(T)$, while $\kappa_{\text{total}}$ represents the sum of the electron and lattice contributions to thermal conductivity $(\kappa_\text{e} + \kappa_\text{l})$. The electrical conductivity, denoted by $\sigma$, can be defined as:  

\[
\sigma = ne\mu .
\tag{6}
\]

Here, $n$ represents the charge carrier concentration, $e$ is the elementary charge, and $\mu$ refers to the charge carrier mobility. Therefore, maximizing electrical conductivity entails maximizing the mobility of the charge carriers, a feat achievable by elevating the temperature of the specimen. The mobility of both holes and electrons, characterized by:  

\[
\mu_h = \frac{e \tau_h}{m_h^*} ,
\tag{7}
\]
and  
\[
\mu_e = \frac{e \tau_e}{m_e^*}, 
\tag{8}
\]
respectively, relies on the factors such as relaxation time $(\tau)$ and effective mass $(m^*)$ of the charge carriers. Consequently, while heavy charge carriers may result in reduced mobility and subsequently in a decreased electrical conductivity, they contribute to enhancing the Seebeck coefficient.

BoltzTraP software was utilized to examine the transport parameters of the investigated substances. Within the semi-classical Boltzmann transport equation paradigm, this computing tool utilizes the constant relaxation time approximation (CRTA) approach. The Seebeck coefficient $(S)$ can be independently evaluated in CRTA without the requirement for extra constraints because the relaxation time $(\tau)$ stays constant. Nonetheless, the relaxation time $(\tau)$ was incorporated in the computations of electrical conductivity $(\sigma)$ and electronic thermal conductivity $(\kappa_\text{e})$, where the computed quantities are expressed as $\sigma / \tau$ and $\kappa_\text{e} / \tau$, respectively. The power factor was assessed in terms of relaxation time, resulting in $(S^2 \sigma / \tau)$, instead of measuring the power factor $(S^2 \sigma)$, as a result of this dependence on relaxation time. 

\begin{figure}[!t]
    \centering
     \includegraphics[width=12cm, height=6cm]{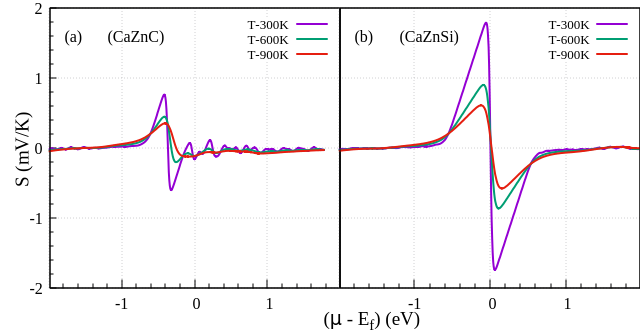} 
    \caption{(Colour online) Change in $S$ with chemical potential at 300, 600, and 900~K.}
    \label{fig:9}
\end{figure}

\begin{figure}[!t]
    \centering
     \includegraphics[width=12cm, height=6cm]{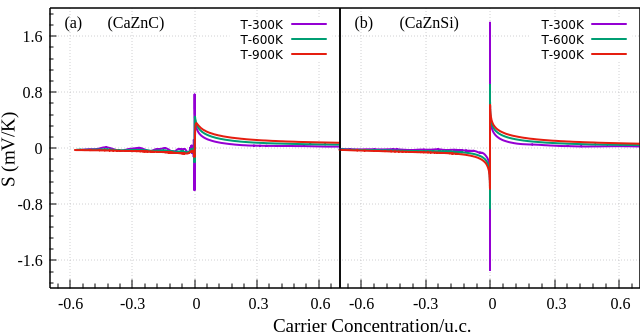} 
    \caption{(Colour online) Change in $S$ with carrier concentration/u.c. at 300, 600, and 900~K.}
    \label{fig:10}
\end{figure}

\begin{figure}[!t]
    \centering
    \includegraphics[width=12cm, height=6cm]{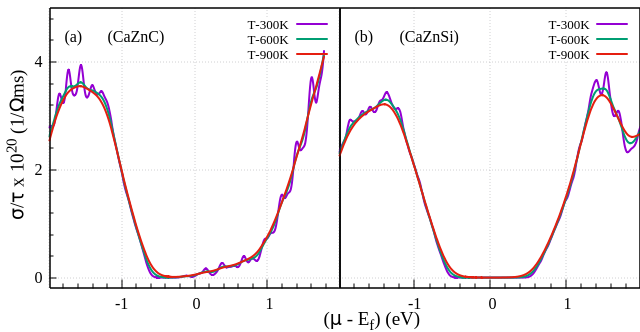} 
    \caption{(Colour online) Change in $\sigma / \tau$ with chemical potential at 300, 600, and 900~K.}
     \label{fig:11}
\end{figure}

\begin{figure}[!t]
    \centering
    \includegraphics[width=12cm, height=6cm]{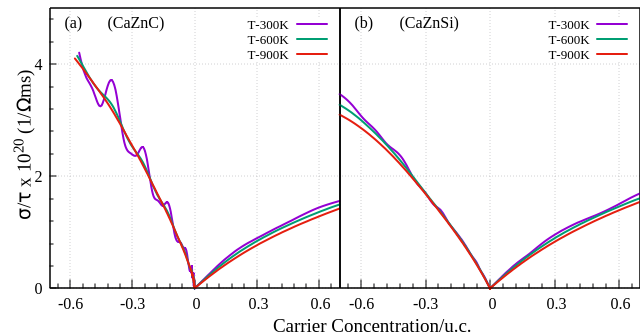} 
    \caption{(Colour online) Change in $\sigma / \tau$ with carrier concentration/u.c. at 300, 600, and 900~K.}
     \label{fig:12}
\end{figure}

\begin{figure}[!t]
\centering
\includegraphics[width=12cm, height=6cm]{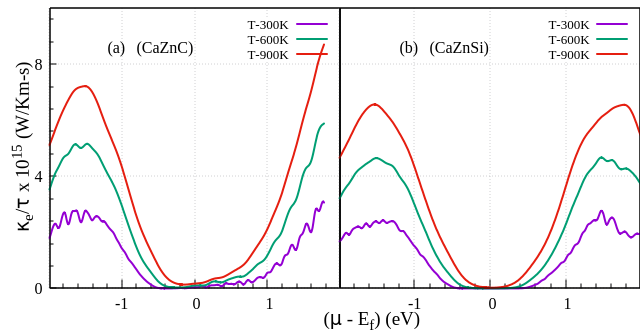}
\caption{(Colour online) Change in \(\kappa_\text{e} / \tau\) with chemical potential at 300, 600, and 900 K.}
\label{fig:13}
\end{figure}

\begin{figure}[!t]
\centering
\includegraphics[width=12cm, height=6cm]{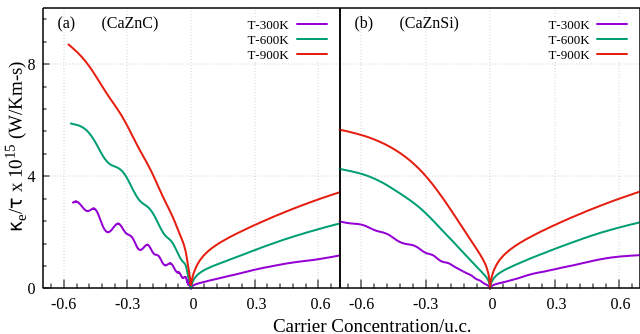}
\caption{(Colour online) Change in \(\kappa_\text{e} / \tau\) with carrier concentration/u.c. at 300, 600, and 900 K.}
\label{fig:14}
\end{figure}

\begin{figure}[!t]
\centering
\includegraphics[width=12cm, height=6cm]{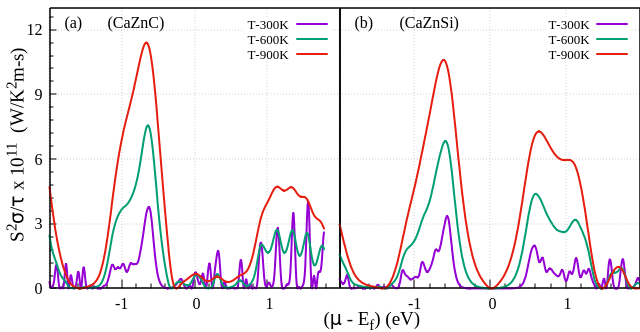}
\caption{(Colour online) Change in PF with chemical potential at 300, 600, and 900 K.}
\label{fig:15}
\end{figure}

\begin{figure}[!t]
\centering
\includegraphics[width=12cm, height=6cm]{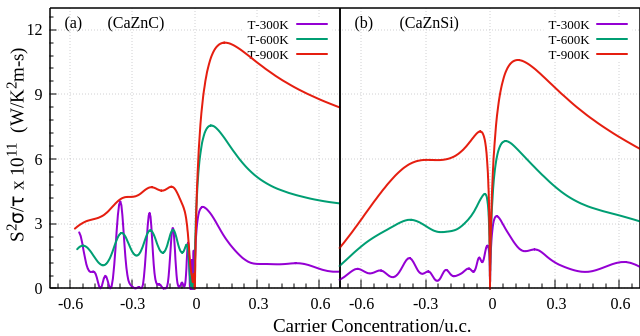}
\caption{(Colour online) Change in PF with carrier concentration/u.c. at 300, 600, and 900 K.}
\label{fig:16}
\end{figure}

\begin{figure}[!t]
\centering
\includegraphics[width=12cm, height=6cm]{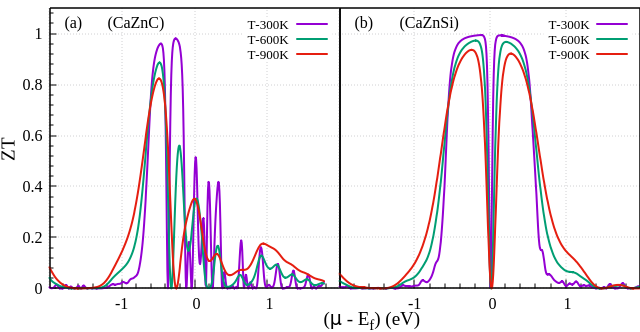}
\caption{(Colour online) Change in ZT with chemical potential at 300, 600, and 900 K.}
\label{fig:17}
\end{figure}

\begin{figure}[!t]
\centering
\includegraphics[width=12cm, height=6cm]{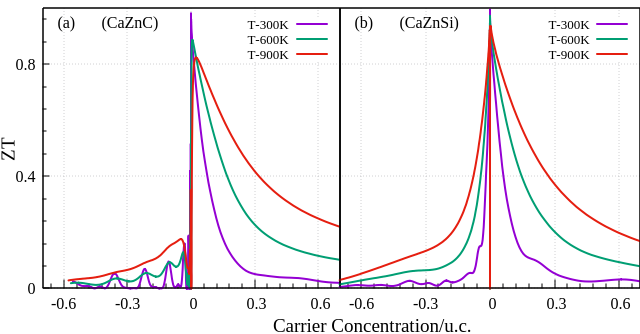}
\caption{(Colour online) Change in ZT with carrier concentration/u.c. at 300, 600, and 900 K.}
\label{fig:18}
\end{figure}

Since the lattice thermal conductivity $(\kappa_\text{l})$ was not included in the BoltzTraP code computations, only the electronic component of thermal conductivity $(\kappa_\text{e})$ was taken into account. As a result, $\kappa_\text{l}$ was calculated using Slack's model~\cite{ref47}. Figures \ref{fig:9}--\ref{fig:18} illustrate how these parameters change with the chemical potential $(\mu - E_f)$ and carrier concentration per unit cell $(n)$ at the designated temperatures. The figures demonstrate that the regions of negative/positive $(\mu - E_f)$ align with $p$-type/$n$-type doping (holes/electrons), whereas for $(n)$, the opposite trend is observed, with negative/positive regions corresponding to the $n$-type/$p$-type area and indicating the concentration of electrons/holes.

The Seebeck coefficient $(S)$ is as an important indicator of a compound's bandstructure, shedding light on the dominant charge carriers within the specimen. To thoroughly understand this behavior, $S$ was computed across varying $(\mu - E_f)$ and $(n)$ at temperatures of 300~K, 600~K, and 900~K, as represented in figures~\ref{fig:9} and~\ref{fig:10}, respectively. From these figures, it is evident that the Seebeck coefficient curve exhibits two distinct peaks, representing the $p$-region and $n$-region of the material. Notably, $S$ tends to decrease with higher temperatures, with its maximum value typically observed at 300~K. The peak Seebeck coefficient $(S)$ values at 300~K and 900~K for both $p$-type and $n$-type regions are depicted in table~\ref{tab:2}. Remarkably, these maxima and minima occur within a narrow range of doping concentrations, indicating that they can be attained with small doping concentrations, a favorable trait for practical applications. Additionally, it is noted that $S$ gradually drops with higher doping concentrations.

Examining the electrical conductivity per unit relaxation time $(\sigma / \tau)$, which is determined as a function of $(\mu - E_f)$ and $(n)$, make available additional data on the electrical characteristics of a substance. These data are shown in figures~\ref{fig:11} and~\ref{fig:12}, correspondingly. The $(\sigma / \tau)$, in contrast to $S$, varies very little with temperature. Higher doping concentrations cause charge carriers to move more freely, as the image shows, which causes electrical conductivity to clearly tend upward. This finding emphasizes how accurate and dependable our computational results are. Table~\ref{tab:2} lists the extreme $(\sigma / \tau)$ values for both materials at 300~K and 900~K. Both the tabularized data and the figure, when thoroughly analyzed, show that the $(\sigma / \tau)$ of both considered compounds is greater in the $n$-type region than in the $p$-type region at all considered temperatures.

\begin{table}[h!]
\centering
\caption{TE characteristics of CaZnC and CaZnSi HH materials.}
\vspace{0.2cm}
\label{tab:2}
\resizebox{\textwidth}{!}{%
\begin{tabular}{l l cc cc cc cc cc}
\hline
\multirow{2}{*}{Materials} & \multirow{2}{*}{Type} & \multicolumn{2}{c}{$S$ (mV/K)} & \multicolumn{2}{c}{$\sigma/\tau$ ($10^{20}$ $(\Omega$\,ms)$^{-1}$)} & \multicolumn{2}{c}{PF/$\tau$ ($10^{11}$ W/K$^{2}$\,m\,s)} & \multicolumn{2}{c}{$N$ (carriers/u.c.)} & \multicolumn{2}{c}{$ZT$} \\
\cline{3-12}
 &  & 300 K & 900 K & 300 K & 900 K & 300 K & 900 K & 300 K & 900 K & 300 K & 900 K \\
\hline
CaZnC  & $p$-type & 0.764  & 0.354  & 3.942 & 3.550 & 3.778 & 11.394 & $1\times10^{-5}$ & 0.02418 & 0.960 & 0.823 \\
       & $n$-type & $-0.157$ & $-0.111$ & 4.203 & 4.099 & 4.035 & 4.713  & $-1.04\times10^{-5}$ & $-0.00122$ & 0.515 & 0.351 \\
CaZnSi & $p$-type & 1.791  & 0.611  & 3.441 & 3.215 & 3.350 & 10.589 & 0 & 0.00191 & 0.993 & 0.936 \\
       & $n$-type & $-1.735$ & $-0.574$ & 3.807 & 3.383 & 1.982 & 7.278  & $-1\times10^{-8}$ & $-0.00075$ & 0.993 & 0.921 \\
\hline
\end{tabular}}
\end{table}

The electronic thermal conductivity per unit relaxation time $(\kappa_\text{e}/\tau)$ plays a crucial role in evaluating the efficacy of TE compounds. As presented in figures~\ref{fig:13} and~\ref{fig:14}, the values of $\kappa_\text{e}/\tau$ have been computed for different $(\mu - E_f)$ and $(n)$ across three selected temperatures. It is imperative for $\kappa_\text{e}/\tau$ to remain low to achieve optimal efficiency in TE applications. As the temperature rises, $\kappa_\text{e}/\tau$ progressively increases, reaching a minimum value at 300~K. These data demonstrate this. The regions in which these materials show the lowest values of $\kappa_\text{e}/\tau$ are in close proximity to the regions where the Seebeck coefficient of these materials reaches its maximum values. Interestingly, for different chemical potentials at a given temperature, $\kappa_\text{e}/\tau$ shows a trend similar to $\sigma/\tau$. Additionally, it can be seen that at any particular temperature, $\kappa_\text{e}/\tau$ also increases with increasing doping concentrations. This pattern highlights the potential suitability of the considered substances for TE devices within the energy and carrier concentration/u.c. range depicted in figures~\ref{fig:13} and~\ref{fig:14}.

By applying the previously described transport properties, the power factor (PF) in relaxation time units \((S^2 \sigma / \tau)\) has been ascertained. The fluctuation in PF with regard to the chemical potential \((\mu - E_f)\) is depicted in figures~\ref{fig:15} and \ref{fig:16}. Interestingly, the power factor reaches its smallest value when \((\mu - E_f)\) approaches the Fermi level. This phenomenon arises because, at this point, the \(\sigma / \tau\) diminishes considerably, while the S approaches zero, leading to minimal TE performance. However, as \((\mu - E_f)\) increases, both S and \(\sigma / \tau\) exhibit a rising trend. Nevertheless, even if \(\sigma / \tau\) rises, there is a drop in S when the chemical potential rises. Two noticeable peaks arise as a result of this behavior, one in the compound's $n$-region and one in its $p$-region.

PF is greater in the $p$-type region than the $n$-type region, as seen in figures \ref{fig:15} and \ref{fig:16}, and its peaks similarly shift towards somewhat higher doping concentrations as the temperature rises. The extreme PF values in the $p$- and $n$-regions close to the Fermi level of these substances at 300~K and 900~K are shown in table~\ref{tab:2}. These values more clearly show the potential TE performance of the considered compounds.

The calculated $ZT$ for these chemicals is based on the discussed transport coefficients. Figures \ref{fig:17} and~\ref{fig:18} show the variation in $ZT$ with regard to \((\mu - E_f)\) and \((n)\), respectively. Table~\ref{tab:2} displays the greatest $ZT$ values in the $p$-region and $n$-region for the investigated materials at 300~K and 900~K. $ZT$ tends to be higher at lower temperatures, as this table shows. The $ZT$ values for the $p$- and $n$-regions of CaZnC and CaZnSi compounds are 0.960 and 0.993, respectively, around the Fermi level, at 300~K. For the $n$-region, these values are 0.523 and 0.936, respectively. These substances are proposed as potential candidates for TE devices because of their high $ZT$ values. 

Table~\ref{tab:2} offers a narrow window of carrier concentrations that can assist the experimentalists in achieving the best carrier concentration/u.c. for the maximum $ZT$ of the materials under study for their $p$- and $n$-type regions at 300 and 900 K.

To compute the lattice thermal conductivity \(\kappa_\text{l}\), we implemented Slack's model \cite{ref47}, as BoltzTraP does not yield this value and treats it as zero in the total thermal conductivity (\(\kappa_{\text{total}}\)). \(\kappa_\text{l}\) can be calculated utilizing Slack's model, as shown in the following equation:

\[
\kappa_\text{l} = B \frac{M \delta \theta_\text{D}^3 n^{1/3}}{T \gamma^2}. 
\tag{9}
\]

Here, 
\[
 B = \frac{(5.72 \times 10^{-8} \times 0.849)}{2(1 - \frac{0.514}{\gamma} + \frac{0.228}{\gamma^2})} 
\]
is a constant derived from the model, \( M \) denotes the average atomic mass, \( \theta_\text{D} \) indicates the acoustic Debye temperature, \( \delta \) signifies the cube root of the mean volume occupied by a single atom, \( \gamma \) signifies the acoustic Gruneisen component, and \( n \) refers to the number of atoms within a unit cell. \( \theta_\text{D} \) serves to characterize the vibrational modes of the substance, while \( \gamma \) indicates the degree of lattice anharmonicity present in the compound.

\begin{figure}[!t]
\centering
\includegraphics[width=0.5\textwidth]{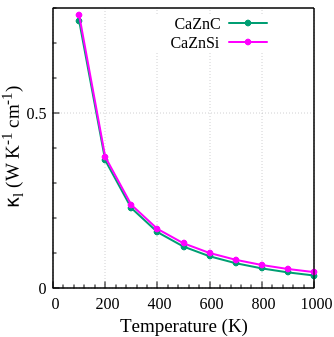} 
\caption{(Colour online) Change in \( \kappa_\text{l} \) with temperature.}
\setlength{\abovecaptionskip}{10pt}
\label{fig:19}
\end{figure}

There has been a lot of work done, using the above equation~\cite{ref48,ref49}, to determine \( \kappa_\text{l} \). For CaZnC and CaZnSi, the \( \kappa_\text{l} \) calculations (at 300~K) give 0.229 and 0.239~W/(K\,cm), respectively. The \( \kappa_\text{l} \) values at 900~K for CaZnC and CaZnSi were 0.046 and 0.055~W/(K\,cm) respectively. This suggests that the \( \kappa_\text{l} \) diminishes with the rising temperature. The general behavior of \( \kappa_\text{l} \) in relation to temperature is shown in figure~\ref{fig:19}.

Numerous factors contribute to improving the $ZT$ of TE compounds. Key contributors include the crystalline phase stability, strain reduction mechanisms, and microstructural modifications, all of which can optimize the charge carrier mobility and thermal transport properties. Additionally, the presence of resonant states induced by dopant atoms plays a crucial role in improving the Seebeck coefficient by altering the electronic DOS. Furthermore, structural ordering, the presence of heavy hole bands, and the composition of doped samples contribute significantly to the TE performance by enhancing the carrier scattering and modifying the bandstructure characteristics. The incorporation of elements with higher atomic mass effectively reduces the lattice thermal conductivity, thereby boosting $ZT$. Lastly, materials with high band degeneracy exhibit an increased number of conductive states, leading to an improved electrical conductivity while maintaining a favorable Seebeck coefficient~\cite{ref50,ref51}. Researchers have proposed various strategies to increase the $ZT$. For example, Poon \cite{ref52} suggested that reducing the strain of HH-compounds by increasing their processing temperature can positively impact their TE characteristics. Chen  et al. \cite{ref53} suggested a number of strategies to improve $ZT$, such as enhancing electrical conductivity through metallic phase nanostructures and tuning VEC. Additionally, improving Seebeck constant through low-energy carrier filtering along with resonant state evolution in the neighborhood of the Fermi level. Ultimately, reducing thermal conductivity by decreasing grain size to the nanoscale, produced by sudden hot pressing of bulk compounds, can further improve the $ZT$.

\subsection{Elastic properties}

Elastic properties serve as an essential foundation for the mechanical response of a material therefore, they provide information on various interesting physical and structural characteristics. These properties contribute to defining the equation of state, which describes the material's response when subjected to stress. Hence, one can make general statements about the intrinsic strength, either ductility or brittleness, of the material depending on the extent of its deformability prior to fracture. The elastic parameters, in addition, are significant for the study of sound propagation through materials and contribute to the acoustic properties and sound velocity. Another important factor is the Debye temperature (\(\theta_\text{D}\)), which is closely linked with lattice vibration, heat capacity, and thermal conductivity. These are used for studying interatomic intertactions under stress and strain, essential for carrying out first-principles studies based on DFT~\cite{ref54}.

In this respect, elastic constants act as critical layers to describe the reaction of the compound to an external force. Through these constant values, critical information is provided to select appropriate compounds for advanced photovoltaic use. Different magnitude-dependent strain configurations of symmetry are computed from the equilibrium structure and optimized for the internal structure. The three independent elastic stiffness coefficients, \(C_{11}\), \(C_{12}\), and \(C_{44}\), were obtained via this computational modelling. The determined values for these constants are presented here in table~\ref{tab:3}, revealing important details regarding the anisotropic elastic behavior and mechanical strength qualities of the studied materials \cite{ref55,ref56}. As to the interpretation of these constants, $C_{11}$ indicates the modulus for uniaxial compression, \(C_{12}\) the modulus for transverse expansion, and \(C_{44}\) the resistance to shear stresses. Large \(C_{11}\) may suggest that CaZnC and CaZnSi are very resistant to compression even when subjected to strong uniaxial pressure. If the following criteria are satisfied, a cubic crystal structure is deemed mechanically robust with respect to elastic deformations: \(C_{11} - C_{12} > 0\), \(C_{44} > 0\), \(C_{11} + 2C_{12} > 0\), and \(C_{11} > 0\)~\cite{ref57}. The values of these constants obtained from the computations showed a perfect fit with the aforementioned equations for all investigated compounds. So, it could be established that these substances have mechanical stability, opening the further study for application potentials in renewable energy technologies.

Further, the VRH approximation, based on stiffening constants, provides a neat way of calculating the mechanical properties of compounds like bulk modulus (\(B\) in GPa) and shear modulus (\(G\) in GPa). A bulk modulus is an essential measure for a compound to resist against a uniform compression, which is measured from its capability to deflect under an applied pressure, while the shear modulus is important to evaluate hardness, as it gives insight into resistance against shear stress. By means of the Voigt and Reuss approximation, the cubic materials had their respective bulk and shear moduli calculated since these approximations present the upper and lower bounds for these elastic properties, respectively \cite{ref58,ref59}:

\[
B_V = B_R = \frac{(C_{11} + 2C_{12})}{3}, \tag{10}
\]

\[
G_V = \frac{(C_{11} - C_{12} + 3C_{44})}{5}, \tag{11}
\]

\[
G_R = \frac{5(C_{11} - C_{12}) C_{44}}{4C_{44} + 3(C_{11} - C_{12})}. \tag{12}
\]

According to Hill's approach, the actual values of the moduli can be determined by taking the arithmetic mean of the moduli computed utilizing Voigt and Reuss approximations~\cite{ref60}:

\[
B = \frac{(B_V + B_R)}{2} \tag{13}
\]
and
\[
G = \frac{(G_V + G_R)}{2} .\tag{14}
\]

This represents a systematic depiction of the mechanical characteristics of compounds, dealing properly with both isotropic and anisotropic mechanical behavior. By keeping both Voigt and Reuss upper and lower approximations, the averaging of their results gives a more complete representation of the mechanical response of the material. Such an approach would increase the extent of evaluation for applications and that is important in understanding the mechanical response of this material under quite a big number of conditions.

The calculated $B$ and $G$ values of the studied materials, derived from the VRH approximation, are carefully filled out in table~\ref{tab:3}. The $B$ defines the resistance of the material against the applied pressure, larger values meaning less compressibility and thus higher resistance to deformation. The $G$ defines the resistance of a compound against plastic deformation and indicates the hardness. The bulk moduli produced via VRH approximation corresponds neatly with those from EOS computations on volume optimization, hence validation of the GGA methods used. The critical $B/G$ ratio limit, as presented by Pugh  et al. \cite{ref61}, acts as a determination of the ductile or brittle nature of the material: a ductile material would have a ratio larger than 1.75, with a higher ratio indicating a better ductility. The computed $B/G$ ratios displayed in table~\ref{tab:3} for CaZnC and CaZnSi are all lower than 1.75, thus attesting to their brittle attribute in ambient conditions. One may further predict their order of brittleness as CaZnSi > CaZnC.

\begin{table}[h]
\centering
\caption{Calculated elastic parameters of CaZnC and CaZnSi at ambient conditions.}
\vspace{0.2cm}
\label{tab:3}
\resizebox{\textwidth}{!}{%
\begin{tabular}{lcccccccccc}
\hline
Compounds & $C_{11}$ (GPa) & $C_{12}$ (GPa) & $C_{44}$ (GPa) & $B$ (GPa) & $G$ (GPa) & $B/G$ & $C'$ & $C''$ & $\eta$ & $Y$ (GPa) \\
\hline
CaZnC & 222.01 & 8.222 & 36.902 & 79.48 & 57.448 & 1.384 & 106.895 & -28.680 & 0.209 & 138.885 \\
      & 241.2$^{a}$ & 5.4$^{a}$ & 31.8$^{a}$ & 84.0$^{a}$ & -- & -- & -- & -- & -- & -- \\
CaZnSi & 74.564 & 9.752 & 28.713 & 31.356 & 30.137 & 1.040 & 32.406 & -18.961 & 0.136 & 68.475 \\
CaMgSi$^{b}$ & 93.49$^{b}$ & 15.92$^{b}$ & 25.407$^{b}$ & 41.78$^{b}$ & 30.12$^{b}$ & 1.39$^{b}$ & -- & -- & -- & 72.84$^{b}$ \\
\hline
\multicolumn{11}{l}{$^{a}$Reference~\cite{ref23}; $^{b}$Reference~\cite{ref39}} \\
\end{tabular}%
}
\end{table}

To further validate the ductile or brittle nature of the considered materials, two very important parameters, the shear constant ($C'$), also called the tetragonal shear modulus, and the Cauchy pressure ($C''$), are calculated. A positive shear constant indicates that a material is dynamically stable, whereas a positive Cauchy pressure points towards metallic bonding, generally associated with ductility. On the contrary, the combined evidence of negative Cauchy pressure, associated with a preponderance of covalent bonding, indicates brittleness~\cite{ref62,ref63}. The expressions that give the $C'$ and $C''$ are as below~\cite{ref64}:

\begin{equation}
C' = \frac{C_{11} - C_{12}}{2}, \tag{15}
\end{equation}

\begin{equation}
C'' = C_{12} - C_{44} .\tag{16}
\end{equation}

These additional parameters provide a deeper insight into the mechanical characteristics of the investigated substances, offering a comprehensive understanding of their response to various external conditions. The determined shear constant ($C'$) and Cauchy pressure ($C''$) are systematically displayed in table~\ref{tab:3}. Notably, the positive values of $C'$ for all compounds confirm their stability against tetragonal distortion, further reinforcing their dynamical stability. Additionally, the negative Cauchy pressure values observed for the investigated materials strongly indicate their brittle nature, aligning with the presence of covalent bonding within the materials.

Poisson's ratio ($\eta$) serves as another crucial parameter for justifying the brittle or ductile nature of a substance. It plays a key role in evaluating the compressibility of a specimen and characterizing its bonding forces. Poisson's ratio of $\eta = 0.5$ implies that the compound is incompressible, with its volume remaining unchanged regardless of deformation. For materials governed by central forces, the Poisson's ratio must lie between 0.25 and 0.5. If $\eta < 0.25$, it suggests the presence of non-central interatomic forces within the material. Furthermore, the nature of bonding within a material can be determined using ($\eta$). A material predominantly exhibits covalent bonding when $\eta \approx 0.1$, whereas a value of $\eta \approx 0.33$ indicates the presence of the metallic bonding. If $0.1 < \eta < 0.33$, the compound possesses a hybrid bonding character, exhibiting both covalent and metallic interactions. Poisson's ratio ($\eta$) can be determined using the formula given below~\cite{ref64}:
\[
\eta = \frac{3B - 2G}{2(3B + G)}. \tag{17}
\]
This parameter expands on our knowledge of the properties of the compounds under various circumstances by providing more information about their mechanical behavior and bonding qualities. The computed values of $\eta$ are presented in Table~\ref{tab:3}. Analyzing these values, it is evident that for the investigated compounds, $\eta$ falls within the range of 0.1 to 0.25, signifying that covalent bonding is the dominant interaction, with a slight metallic character. This finding is in strong agreement with prior studies on their electronic parameters, further reinforcing the covalent nature of these compounds.

Young's modulus ($Y$ in GPa) serves as an essential characteristic for assessing a compound's hardness, describing its stiffness. It provides insights into how the substance responds to longitudinal strain along its edges, with greater Young's modulus values indicating greater stiffness. Young's modulus is computed using the formula:
\[
Y = \frac{9BG}{3B + G}. \tag{18}
\]
The computed values of Young's modulus are documented in table~\ref{tab:3}. Analysis of this table reveals the order of stiffness for these materials as CaZnC > CaZnSi.

Considering the influence of crystal structure, different crystal structures exhibit varying degrees of anisotropy. Hence, the Zener anisotropy index ($A_Z$) is adopted for cubic structures. An elastically isotropic compound has an $A_Z$ value of 1, while the values other than 1 indicate anisotropy. The Zener anisotropic factor can be calculated by using the elastic constants ($C_{11}$, $C_{12}$, and $C_{44}$) as given below:
\[
A_Z = \frac{2C_{44}}{C_{11} - C_{12}}. \tag{19}
\]
The Zener anisotropic factor ($A_Z$) for the examined compounds is presented in table~\ref{tab:4}. The observed $A_Z$ values are below unity, signifying that the investigated compounds exhibit significant elastic anisotropy. This pronounced anisotropic behavior suggests a higher susceptibility to microcrack formation, which could influence their mechanical stability and durability under external stress.

Kleinman introduced a significant parameter, known as the Kleinman parameter ($\zeta$), to measure the internal strain within a material. This parameter is essential for describing the bending and stretching behavior of bonds within a solid material. When $\zeta$ equals 0, it signifies a limitation on the bond bending, whereas $\zeta$ approaching 1 indicates dominance in bond stretching. The Kleinman parameter can be derived from the elastic constants using the following formula~\cite{ref65,ref66}:
\[
\zeta = \frac{C_{11} + 8C_{12}}{7C_{11} + 2C_{12}}. \tag{20}
\]
The computed $\zeta$ values are systematically shown in table~\ref{tab:4}. A notable observation is that for all the investigated compounds, $\zeta$ tends towards zero, signifying that the bond bending effects predominate over the bond stretching. This characteristic strongly correlates with high brittleness, reaffirming the intrinsic mechanical fragility of these materials.

Lame's coefficients, namely the first coefficient ($\lambda$) and the second coefficient ($\mu$), serve as fundamental mechanical characteristics derived from stress-strain relationship. These coefficients provide crucial insights into the elastic behavior of materials, particularly in describing their response to mechanical deformation. In the case of homogeneous isotropic materials, their elastic properties can be effectively characterized using Hooke's law in three dimensions, which establishes a mathematical relationship between stress and strain. The formulation of Hooke's law for such systems is given by~\cite{ref67}:
\[
\sigma = 2\mu \alpha + \lambda \, \text{tr}(\alpha) I .\tag{21}
\]

Here, $\sigma$ denotes stress, $\alpha$ is the strain tensor, and $I$ signifies the identity matrix. These coefficients are essential in defining the elastic moduli of homogeneous isotropic substances and are related to various material properties. Physically, the first Lame constant ($\lambda$) is linked to the material's compressibility, while the second constant ($\mu$) reflects its shear stiffness~\cite{ref68}. These coefficients can be determined using Young's modulus ($Y$) and Poisson's ratio ($\eta$) as follows:
\[
\lambda = \frac{Y \eta}{(1 + \eta)(1 - 2 \eta)}, \tag{22}
\]
\[
\mu = \frac{Y}{2(1 + \eta)}. \tag{23}
\]
The computed $\lambda$ and $\mu$ values are systematically presented in table~\ref{tab:4}. A thorough examination of these values reveals that both $\lambda$ and $\mu$ exhibit consistently low magnitudes for the investigated materials. This observation suggests that these materials possess low compressibility and limited resistance to shear deformation. Such findings are in excellent agreement with other derived elastic parameters, reinforcing the reliability and coherence of the obtained mechanical property data.

\begin{table}[!t]
\centering
\caption{(Colour online) Computed other elastic parameters of CaZnC and CaZnSi at ambient conditions.}
\vspace{0.2cm}
\label{tab:4}
\resizebox{\textwidth}{!}{%
\begin{tabular}{lcccccccc}
\hline
Compounds & $A_Z$ & $\zeta$ & $\lambda$ & $\mu$ & $v_l$ (m/s) & $v_t$ (m/s) & $v_m$ (m/s) & $\theta_\text{D}$ (K) \\
\hline
CaZnC & 0.345 & 0.183 & 41.187 & 57.448 & 12298.895 & 7461.491 & 8245.184 & 616.774 \\
CaZnSi & 0.886 & 0.282 & 11.265 & 30.137 & 9358.606 & 6074.228 & 6662.511 & 441.744 \\
\hline
\end{tabular}%
}
\end{table}

In addition to elastic constants, the Debye temperature ($\theta_\text{D}$) stands out as an important thermal characteristic that is closely associated with several physical properties, including the melting point and specific heat of the substances. $\theta_\text{D}$ can be computed using the average sound velocity ($v_m$) with the following formula~\cite{ref69}:
\[
\theta_\text{D} = \frac{h}{k} \left[ \frac{3n}{4\piup} \left( \frac{N_A \rho}{M} \right) \right]^{\frac{1}{3}} v_m. \tag{24}
\]
Here, $h$ and $k$ correspond to Planck's constant and Boltzmann constant, respectively, while $n$ signifies the number of atoms per formula unit, and $N_A$ represents Avogadro's number. Additionally, $\rho$ signifies the material's density, and $M$ refers to its molecular mass. A higher sound velocity or elevated $\theta_\text{D}$ typically indicates a stronger interatomic bonding, reflecting the material's robust mechanical integrity and enhanced lattice stability.

For polycrystalline materials, the average sound velocity ($v_m$) is determined using the transverse ($v_t$) and longitudinal ($v_l$) sound velocities. These velocities can be expressed in terms of elastic constants ($B$ and $G$) through Navier's equation~\cite{ref70}:
\[
v_m = \left[ \frac{1}{3} \left( \frac{2}{v_t^3} + \frac{1}{v_l^3} \right) \right]^{-\frac{1}{3}}. \tag{25}
\]

The transverse and longitudinal sound velocities can be calculated as follows~\cite{ref71}:
\[
v_t = \sqrt{\frac{G}{\rho}}, \tag{26}
\]
\[
v_l = \sqrt{\frac{3B + 4G}{3\rho}}. \tag{27}
\]

The values of $v_t$, $v_l$, $v_m$, and $\theta_\text{D}$ gained from the aforementioned equations are presented in table~\ref{tab:4}. Analysis of these outcomes allows for an understanding of the bond strength order among the compounds as CaZnC > CaZnSi. These calculations offer valuable insights into the mechanical behavior of materials, providing a deeper understanding of their behavior under different conditions.

\subsection{Thermodynamic parameters}
\begin{figure}[!t]
	\centering
	\includegraphics[width=14cm, height=10cm]{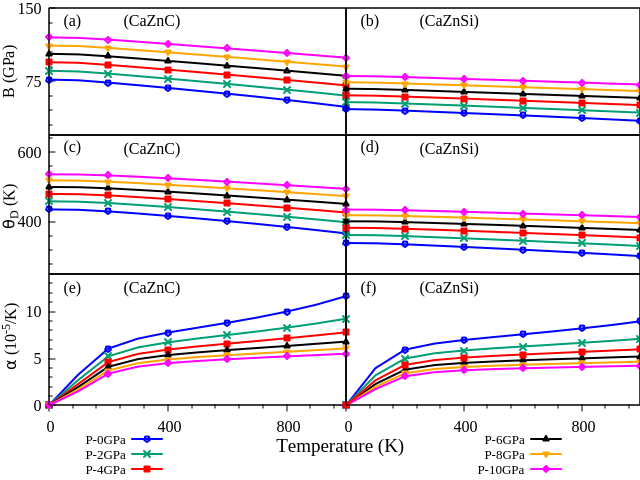}
	\caption{(Colour online) Change in $B$, $\theta_\text{D}$, and $\alpha$ with temperature at a constant pressure, increasing in 2~GPa increments.}
	\label{fig:20}
\end{figure}

The thermodynamic parameters of the considered materials were computed utilizing the Gibbs2 program~\cite{ref37}. In this method, the total energy values, $E(V)$, as a function of volume using several primitive cells were inserted into classical thermodynamic equations to yield the macroscopic thermal properties over a range of temperatures and pressures~\cite{ref72,ref73}. The computations were conducted over a temperature range of 0~K -- 1000~K, while considering pressure effects within the range of 0~GPa $-$ 10~GPa. Implementing the Quasi-harmonic model enabled the inclusion of pressure effect in evaluating the thermodynamic properties. This model allowed for the investigation of changes in the bulk modulus with temperature variations across different pressure levels.

As depicted in figures~\ref{fig:20}(a) and~\ref{fig:20}(b), it was observed that at specific pressures, the bulk modulus of CaZnC and CaZnSi exhibited a gradual decline with increasing temperatures, while at certain temperatures, it experienced a significant rise with increasing pressure. Two fundamental parameters considered in the quasi-harmonic Debye model, the Gruneisen parameter ($\gamma$) and the acoustic Debye temperature ($\theta_\text{D}$), play essential roles in characterizing the thermal properties of materials. The Gruneisen parameter ($\gamma$) is a crucial thermodynamic quantity, expressed through both macroscopic and microscopic properties. From a macroscopic perspective, it is closely associated with the key thermal attributes, including the coefficient thermal expansion (CTE), heat capacity and isothermal bulk modulus. On the other hand, from a microscopic perspective, it correlates with the vibrational frequencies of atoms within the specimen~\cite{ref74}. For the investigated compounds, the values of $\gamma$ are represented in table~\ref{tab:5}.

\begin{figure}[!t]
	\centering
	\includegraphics[width=14cm, height=10cm]{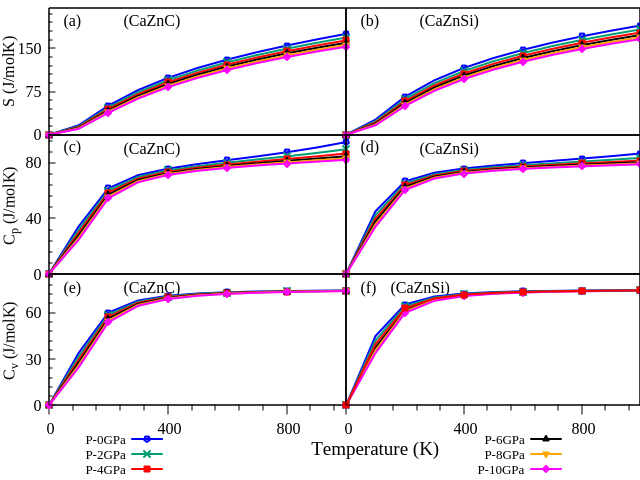}
	\caption{(Colour online) Changes in $S$, $C_p$, and $C_v$ with temperature at a constant pressure, increasing in steps of 2 GPa.}
	\label{fig:21}
\end{figure}

Figures~\ref{fig:20}(c) and \ref{fig:20}(d) illustrate the variation in $\theta_\text{D}$ of CaZnC and CaZnSi as a function of temperature, providing further insights into the thermal behavior of the investigated materials. At a given pressure, the behavior of $\theta_\text{D}$ with respect to temperature exhibits a notable decrease, following a nearly linear trend. Conversely, at a fixed temperature, $\theta_\text{D}$ demonstrates an increase with the rising pressure. This inverse relationship between pressure and temperature significantly influences the Debye temperature, indicating a substantial impact of both factors on the vibration frequency of particles within the studied compounds. Figures~\ref{fig:20}(e) and \ref{fig:20}(f) depict the temperature-dependent behavior of the CTE under varying pressure conditions. The analysis reveals a rapid surge in CTE within the temperature 0~K $-$ 200~K range, indicative of heightened lattice vibrations at lower temperatures. When the temperature rises beyond this point, the increase slows down. This indicates that thermal expansion is stabilizing. Beyond 300~K, the CTE shows an almost linear increase. This is more evident in high pressure conditions. The slope is slight, indicating steady thermal response of the material.

\begin{table}[!t]
\centering
\caption{Thermodynamic characteristics of CaZnC and CaZnSi HH materials at 300 K.}
\vspace{0.2cm}
\label{tab:5}
\resizebox{\textwidth}{!}{%
\begin{tabular}{lccccccc}
\hline
Compounds & $B$ (GPa) & $\theta_\text{D}$ (K) & $\alpha$ ($10^{-5}$ K) & $\gamma$ & $S$ (J/mol\,K) & $C_p$ (J/mol\,K) & $C_v$ (J/mol\,K) \\
\hline
CaZnC & 70.824 & 423.95 & 7.114 & 2.166 & 77.502 & 70.997 & 67.86 \\
CaZnSi & 43.606 & 331.77 & 6.583 & 1.702 & 94.48 & 72.814 & 70.447 \\
\hline
\end{tabular}%
}
\end{table}

Entropy ($S$) is a crucial factor. It starts from absolute zero temperature and increases progressively with temperature. This behavior is shown in figures~\ref{fig:21}(a) and \ref{fig:21}(b). When pressure increases, $S$ decreases slightly. This reduction is modest in magnitude. Entropy measures randomness in the system. Its increase with temperatures indicates that the material becomes more disordered and random~\cite{ref75}. This trend is due to the generation of a bigger thermal energy as temperature rises. The added energy causes stronger vibrations within the system. The rise in entropy comes from these vibrational motions. These motions are anharmonic in nature. To understand the vibrational traits of the substances, its molar heat capacities must be considered. These are measured at constant pressure ($C_p$) and at constant volume ($C_v$).

Figures~\ref{fig:21}(c)--\ref{fig:21}(d) and \ref{fig:21}(e)--\ref{fig:21}(f) show how $C_p$ and $C_v$ changes with temperature and pressure. $C_p$ serves as a pivotal thermodynamic characteristic of a substance because it describes how the substance takes on energy and releases energy as the temperature changes while pressure remains constant. $C_p$ informs the user about how energy is held and transferred. By contrast, $C_v$ describes how the substance takes on energy and releases energy as the temperature changes while the volume remains constant. Specifically, $C_v$ informs the user about the capability of the substances to store energy internally and transfer energy for temperature regulation. Examination of these figures reveals that, for the materials studied, $C_p$ exceeds $C_v$ at any given temperature and pressure. This discrepancy arises because, at a constant pressure, the substance can absorb heat to augment its internal energy and perform work by expanding against external pressure, whereas $C_v$ solely accounts for the heat needed to elevate the internal energy of the substance without any contribution from the work done against external pressure due to no volume alteration. The figures show that heating capacity rises rapidly in the temperature range $0-300$~K, then increases gradually at higher temperatures, nearing the Dulong-Petit limit. This trend is typical of all solids at high temperatures~\cite{ref76}. Table~\ref{tab:5} lists the calculated thermodynamic parameters for all the studied compounds, providing core insights into their thermal and thermodynamic behavior.

\section{Conclusion}

This detailed investigation explores the diverse characteristics of CaZnC and CaZnSi HH materials. Lattice constants were determined as 5.739 \AA\ for CaZnC and 6.475 \AA\ for CaZnSi, with CaZnSi exhibiting greater compressibility. Both materials manifest semiconductor behavior, with bandgaps measured at 1.186~eV for CaZnC and 1.067~eV for CaZnSi. Optical analysis highlights their potential in photovoltaic applications, with CaZnC suited for the UV region and CaZnSi for the visible region. Additionally, they offer UV shielding capabilities, particularly in high UV environments. Thorough examination of TE properties reveals promising $ZT$ of 0.960 for CaZnC and 0.993 for CaZnSi at 300~K. Elastic constants and mechanical parameters indicate brittleness and covalent bonding. Moreover, thermodynamic properties such as entropy and molar heat capacities ($C_p$ and $C_v$) provide insights into their thermal behavior. These findings underscore the versatility of the materials for various technological applications, guiding future research in materials science and in engineering.

\section*{Acknowledgement}
The authors are thankful to the Centre for Research, Instrumentation \& Development (CRID), Poornima University, Jaipur for providing the necessary facilities.

\ukrainianpart

\title{Першопринципне дослідження механічних та функціональних властивостей нових напівгейслерових матеріалів CaZnC та CaZnSi} 
\author {
	П. К. Камлеш\refaddr{label1}, %
	У. К. Гупта\refaddr{label2}, %
	С. Верма\refaddr{label1}, %
	М. Рані\refaddr{label3}, %
	І. Туал\refaddr{label4}, %
	А. С. Верма\refaddr{label5,label6}}

\addresses{
	\addr{label1} Фізичний фукультет університету Пурніма, Джайпур 303905, Раджастан, Індія
	\addr{label2} Школа електротехніки та енергетики, Міжнародний інженерний коледж Ананда, Канота, Джайпур 303012, Раджастан, Індія
	\addr{label3} Фізичний фукультет університету Моханлала Сукхадії, Удайпур 313001, Раджастан, Індія
	\addr{label4} Лабораторія фізики твердого тіла, Університет Сіді Мохамеда Бен Абделли, факультет природничих наук, BP 1796, Фес, Марокко
	\addr{label5} Фізичний фукультет, Інженерно-технологічна школа Ананда, Університет Шарда, Кітхем, Агра 282007, Індія
	\addr{label6} Університетський центр досліджень та розробок, кафедра фізики, Чандігархський університет, Мохалі, Пенджаб 140413, Індія}

\makeukrtitle

\begin{abstract}
	\tolerance=3000%
	У даній роботі представлено результати DFT-розрахунків за допомогою методу FP-LAPW+lo в програмному забезпеченні WIEN2k для отримання інформації про структурні, термоелектричні та оптоелектронні ха\-рак\-те\-рис\-ти\-ки матеріалів CaZnC та CaZnSi. Структурну оптимізацію виконано з використанням функціоналу PBE-GGA, тоді як решту характеристик отримано за допомогою підходу PBE-GGA + TB-mBJ. Термо\-елект\-рич\-нi параметри були оцінені з використанням пакету BoltzTraP. Пружні сталі та інші механічні параметри були обчислені за допомогою коду ELAST у програмному забезпеченні WIEN2k, тоді як термодинамічні характеристики були оцінені за допомогою програми Gibbs2. Результати показують кореляцію між атомним складом та розмірами кристалічної ґратки; при цьому було виявлено, що CaZnC має пряму ($\Gamma$--$\Gamma$) заборонену зону $1.186$~еВ, тоді як CaZnSi має непряму ($\Gamma$--$X$) заборонену зону $1.067$~еВ. Оптичні дослідження сполук прогнозують потенційне застосування для фотоелектричних систем, тоді як термоелектричні результати виявляють оптимізовані коефіцієнти потужності та значення добротності для конверсії енергії. Параметри пружності CaZnC та CaZnSi вказують на стійкість та крихкість матеріалу. Зрештою, термодинамічні оцінки дають інформацію про тепловий механізм та невпорядкованість матеріалів. Як результат, дана дослідницька робота забезпечує значний прогрес у розумінні основ цих сполук та висвітлює їх перспективне застосування в технологіях відновлюваної енергетики.

	\keywords оптичні властивості, структурна стійкість, термодинамічна стійкість, напівгейслерові матеріали, термоелектричні матеріали
\end{abstract}

\end{document}